\documentclass[ reprint, amsmath,amssymb,aps,superscriptaddress]{revtex4-2}
\usepackage{color}
\usepackage{verbatim}
\usepackage{lipsum}
\usepackage{hyphenat}
\usepackage{dirtytalk}
\usepackage{graphicx}% Include figure files
\usepackage{dcolumn}% Align table columns on decimal point
\usepackage{bm}% bold math
\usepackage{tabularx}
\usepackage{textcomp}
\usepackage{hyperref}% add hypertext capabilities
\usepackage{mathtools}
\usepackage{subfigure}
\usepackage{tabularx}
\usepackage{etoolbox}
\usepackage{colortbl}
\usepackage{multirow}
\usepackage{amsmath}
\usepackage{booktabs, makecell}
\usepackage{esint}
\usepackage{xcolor}

\begin{document}
	\title{Dielectric Properties of Single Crystal Calcium Tungstate}
	
    \author{Elrina Hartman}
	\affiliation{Quantum Technologies and Dark Matter Labs, Department of Physics, University of Western Australia, 35 Stirling Highway, Crawley, WA 6009, Australia.}
	\email{To whom correspondence should be addressed: \\ michael.tobar@uwa.edu.au \\ elrina.hartman@research.uwa.edu.au}
	\author{Michael E Tobar}
	\affiliation{Quantum Technologies and Dark Matter Labs, Department of Physics, University of Western Australia, 35 Stirling Highway, Crawley, WA 6009, Australia.}\author{Ben T McAllister}%
	\affiliation{Quantum Technologies and Dark Matter Labs, Department of Physics, University of Western Australia, 35 Stirling Highway, Crawley, WA 6009, Australia.}
    \affiliation{School of Science, Computing and Emerging Technologies, Swinburne University of Technology, Melbourne, VIC 3122, Australia.}
	\author{Jeremy Bourhill}
	\affiliation{Quantum Technologies and Dark Matter Labs, Department of Physics, University of Western Australia, 35 Stirling Highway, Crawley, WA 6009, Australia.}
	\author{Andreas Erb}
	\affiliation{Physik Department Kristalllabor Technische Universität München James Franckstr. D-85748 Garching, Germany.}
    \author{Maxim Goryachev}
	\affiliation{Quantum Technologies and Dark Matter Labs, Department of Physics, University of Western Australia, 35 Stirling Highway, Crawley, WA 6009, Australia.}
    \date{\today}
\begin{abstract}		
This investigation employed microwave whispering gallery mode (WGM) analysis to characterise the dielectric properties of a cylindrical, single-crystal sample of calcium tungstate (CaWO$_4$). Through investigation of quasi-transverse magnetic and quasi-transverse electric mode families, we can assess loss mechanisms and relative permittivity from room temperature to cryogenic conditions. We report the biaxial permittivity values of $\varepsilon^{295\, K}_{||} = 9.029 \pm 0.009$ and $\varepsilon^{295\, K}_{\perp} = 10.761 \pm 0.011$ at room-temperature, and $\varepsilon^{4\, K}_{||} = 8.794 \pm 0.009$ and $\varepsilon^{4\, K}_{\perp} = 10.440 \pm 0.010$ at liquid helium temperature. Components are denoted with respect to the c-axis of the crystal unit cell. The parallel component agrees well with the published literature at MHz frequencies; however, the measured perpendicular component is $4.8$\% lower. The WGM technique offers greater precision, with accuracy limited primarily by the uncertainty in the crystal's dimensions. WGMs also serve as sensitive probes of lattice dynamics, enabling monitoring of temperature-dependent loss mechanisms. At room temperature, the measured loss tangents were $\tan\delta_{||}^{295\,\mathrm{K}} = (4.1 \pm 1.4) \times 10^{-5}$ and $\tan\delta_{\perp}^{295\,\mathrm{K}} = (3.64 \pm 0.92) \times 10^{-5}$. Upon cooling to 4 K, the loss tangents improved by approximately two orders of magnitude, reaching $\tan\delta_{||}^{4\,\mathrm{K}} = (1.56 \pm 0.52) \times 10^{-7}$ and $\tan\delta_{\perp}^{4\,\mathrm{K}} = (2.05 \pm 0.79) \times 10^{-7}$. These cryogenic values are higher than those reported in prior studies, likely due to a magnetic loss channel associated with an unidentified paramagnetic spin ensemble. These findings have implications for the use of CaWO$_4$ in applications such as spin-based quantum systems and cryogenic bolometry, highlighting the potential of WGMs for novel sensing applications.
	\end{abstract}
	\maketitle

\section{Introduction}
Calcium tungstate (CaWO$_{4}$), otherwise known as scheelite, is a scintillating dielectric crystal with low nuclear spin, making it uniquely favourable both in hosting spin qubits for quantum information storage and manipulation\cite{Bertaina2007,Abobeih2018,PhaseMem,PhysRevLett.103.226402}, and in bolometric sensing of rare events\cite{ZDESENKO2005249,ANGLOHER2005325,PhysRevC.70.064606}. As such, much legacy work exists for the characterisation of its thermal properties\cite{Senyshyn2006ThermalEA,ThermalProp,MGluyas1973,NAJAFVANDZADEH2020101089}, scintillation\cite{10.1063/1.4934741,MIKHAILIK2004585,FERRE2023112323}, microwave to optical conversion\cite{CHANELIERE2024120647}, and spectroscopy of rare-earth dopants \cite{ce3+,Wortman,PhysRevB.95.064427,10.1063/1.1674124,nd3+,10.1063/1.1840400,10.1063/1.1669893}. This investigation aimed to contribute to the amassed body of work regarding CaWO$_{4}$ by characterising its dielectric properties over a broad range of temperatures. A high-purity single-crystal sample was grown via the Czochralski method at the Technical University of Munich\cite{C2CE26554K}.  We present the results of studying the temperature dependence of the dielectric constants and associated loss mechanisms in this sample of CaWO$_4$. The dielectric constants for CaWO$_{4}$ have previously been found at room-temperature at $1.59$ kHz by W. S. Brower and P. Fang \cite{BowerandFang,DCofScheelite}, a result later corroborated by J. Thorp and E. Ammar \cite{Thorp1975}, at $1-40$ MHz. This investigation employed microwave whispering gallery mode (WGM) analysis between $2$ to $13$ GHz, consistent with the methodology established by Krupka et. al.\cite{JerzyKrupka_1999}. This study was extended to cryogenic conditions at which, to the authors' knowledge, the dielectric constants have not been measured before. The paper is structured as follows: Section I will conclude with a brief overview of relevant WGM theory. Section II will describe the characterisation procedure to establish WGM identities, which preceded the determination of the dielectric constants at room-temperature. Section III discusses the low-temperature characterisation, outlining the experimental setup and presenting the temperature coefficients of the dielectric permittivity. Section IV highlights the loss mechanisms in both room-temperature and cryogenic conditions. The conclusions of this work are discussed in Section V.\\\\
We undertook experiments and simulations to understand the population of modes in the cylindrical CaWO$_{4}$ sample. WGMs are classified according to the dominant transverse component of the electromagnetic field. Quasi-transverse magnetic and quasi-transverse electric WGMs are designated as WGH$_{m,n,p}$ and WGE$_{m,n,p}$, respectively, where \textit{m} is the azimuthal mode number indicating the number of antinodes in $180$\textdegree \, in the $\phi$-direction, ($m\geq 1$),  \textit{n} is the number of radial (\textit{r}) antinodes ($n\geq 1$), and $p$ is the number of axial (\textit{z}) antinodes ($p\geq 1$). The azimuthal component exhibits total internal reflection about the perimeter of a cylindrical resonator for an integer multiple of the wavelengths ($\lambda$), satisfying the approximate resonant condition, $ m \lambda \approx 2 \pi r_{eff} \sqrt{\varepsilon_r \mu_r}$. Here, $r_{eff}$ is the effective propagation radius of the mode set by the average of the radius of the cylinder and inner radial caustic of the internally reflected rays \cite{Jiao:90}, $\varepsilon_r$ is the relative permittivity, and $\mu_r = 1$ is the relative permeability of the dielectric material. This approximation becomes increasingly accurate in the large-\textit{m} limit. In this regime, the free spectral range (FSR) may be written as,
\begin{equation}
\Delta f=\frac{c}{2\pi r_{eff}\sqrt{\varepsilon_r\mu_r}}, 
\label{eq:fsr}
\end{equation}
and the modal structure asymptotically approaches that of a pure WGM. Eq. \ref{eq:fsr} conveys the relationship between the resonant frequencies, the geometry, and the dielectric properties of the crystal.
\begin{figure}[!t]
	\centering
\includegraphics[width=\columnwidth]{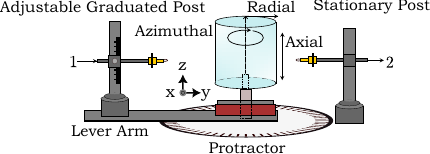}
			\caption{Experimental setup for electromagnetic mode identification in the cylindrical crystal. The radiative fields around the unshielded crystal were measured in transmission using coaxial loop probes (1 and 2) along the axial and azimuthal directions to confirm WGM mode numbers and polarisation. The orientation of the probes and their distance from the crystal surface dictated the strength of the coupling to microwave resonances of distinct mode families.}
			\label{fig:nc_setup}
\end{figure}
\begin{figure}[t!]	\includegraphics[width=\columnwidth]{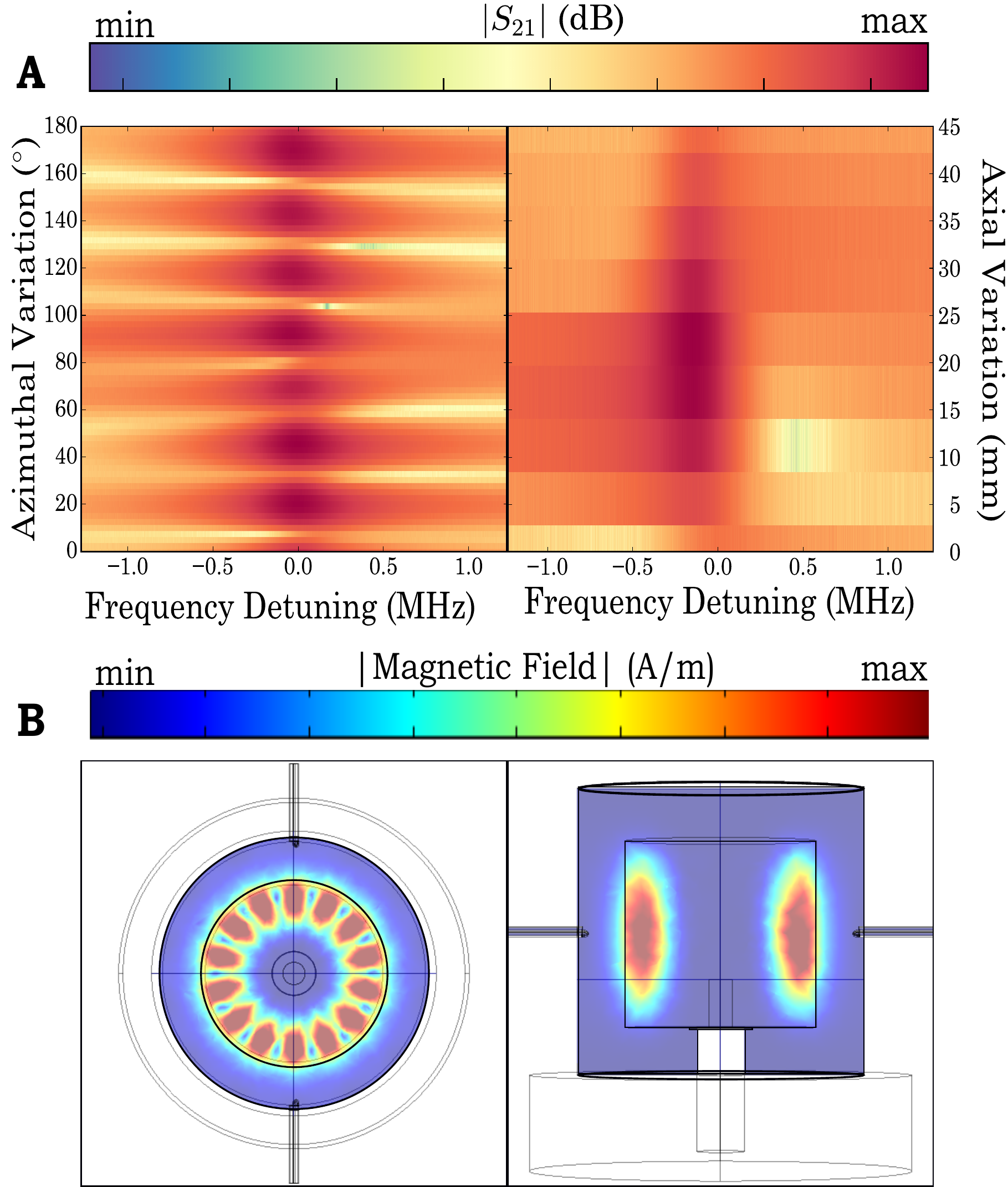}
			\caption{Measured and simulated electromagnetic field variations for the WGE$_{7,1,1}$ mode ($f_{0}=7.965$ GHz). \textbf{A} Colour density plot of the measured transmission as a function of probe position near the surface of the unshielded dielectric, in the azimuthal direction (left), and the axial direction (right). \textbf{B} The magnetic field distribution within the CaWO$_4$ cylindrical dielectric crystal, as simulated by COMSOL.}
			\label{fig:hz_field}
\end{figure}
\begin{figure}[!h]
	\centerline{\begin{minipage}{\columnwidth}
			\centering
\includegraphics[width=0.792\columnwidth]{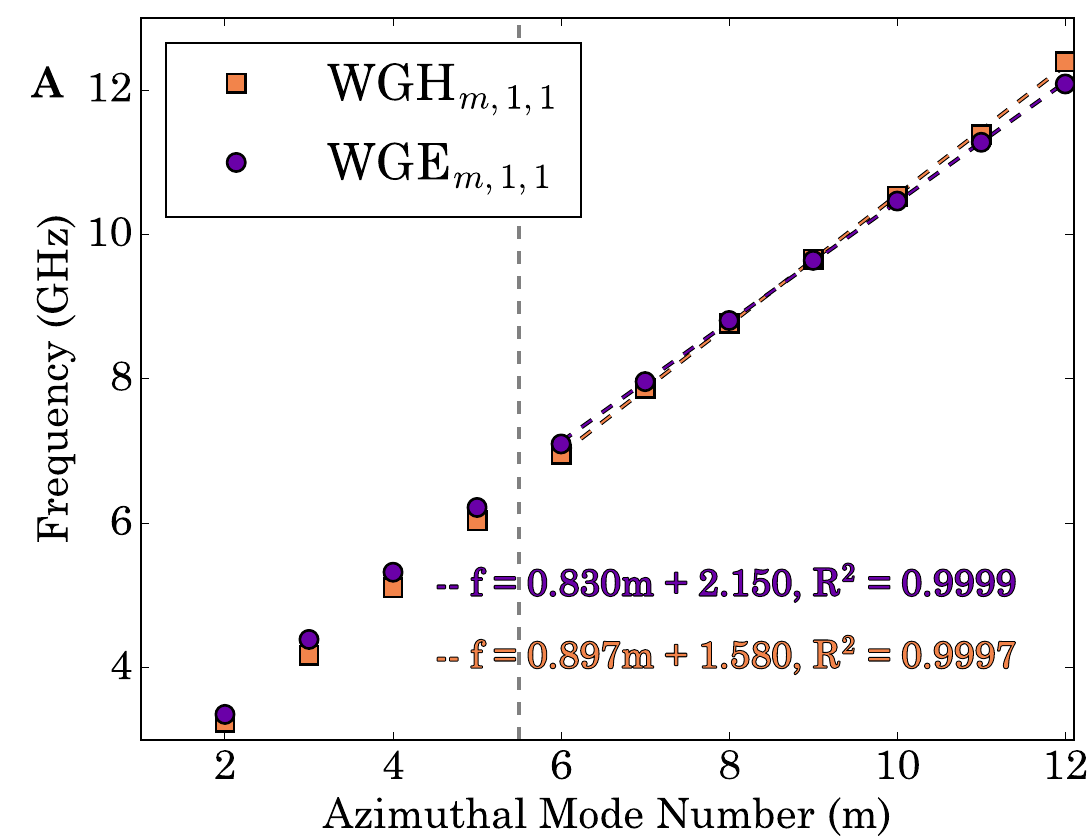}
\includegraphics[width=0.792\columnwidth]{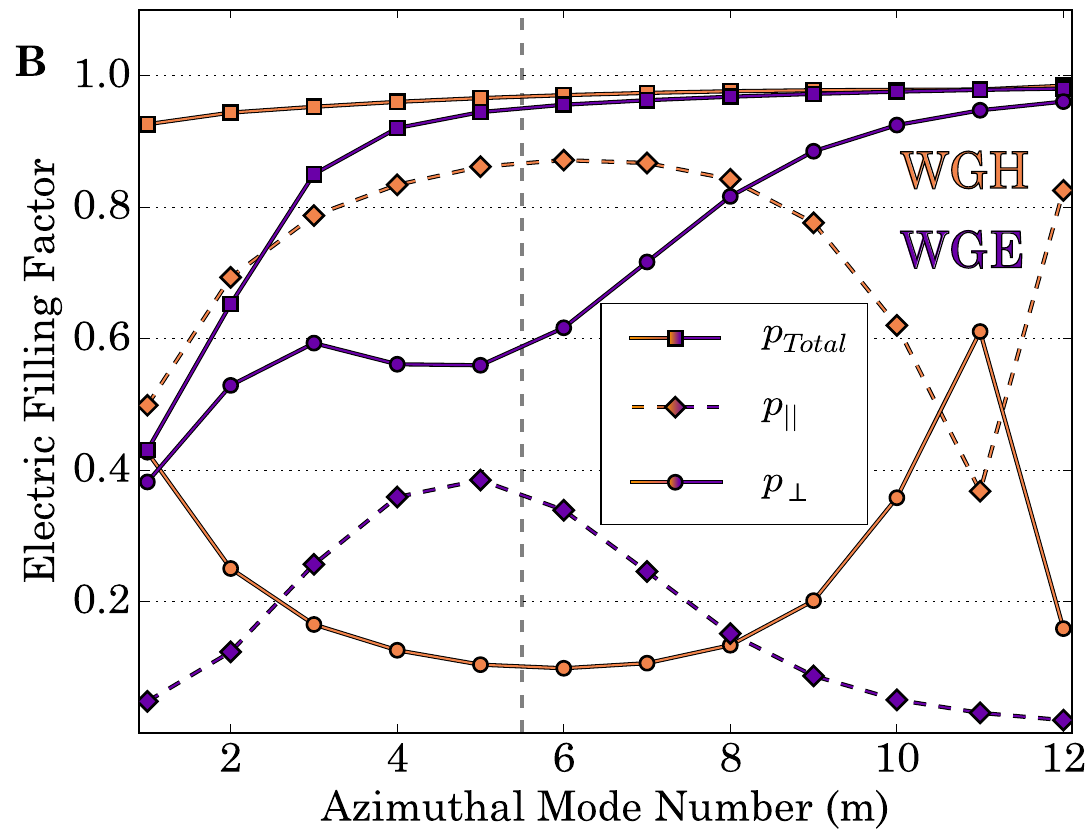}
			\caption{Simulated room-temperature (295 K) characteristics of the fundamental whispering gallery mode families, WGH$_{m,1,1}$ (orange) and WGE$_{m,1,1}$ (purple), plotted as a function of azimuthal mode number, $m$. The vertical, grey, dashed line indicates the threshold above which microwave resonances were considered sufficiently WGM-like for use in subsequent analysis. \textbf{A} Fundamental mode family frequencies. The linear regression is fitted for $m\geq6$, where $m$ is large enough to present a linear frequency dependence. The fit gives an FSR of 0.830 GHz for WGE$_{m,1,1}$ modes and 0.897 GHz for WGH$_{m,1,1}$ modes. Using Eq. \ref{eq:fsr} and assuming the modes are pure TM and TE, we obtain an average radius of propagation of $r_{eff}=17.9$ and $17.8$ mm, respectively. \textbf{B} Electric filling factors are theoretically predicted and employed for the calculation of the temperature coefficients.}
			\label{fig:m_vs}
	\end{minipage}}
\end{figure}
The scheelite structure belongs to the space group C$_{4k}^{6}-I4_{1/a}$ \cite{PhysRevB.57.12738}. The symmetry afforded by the tetragonal unit cell structure allows us to simplify the Cartesian coordinate system (x, y, z) in terms of the c-axis, with z $||$ c  and a = b $\perp$ z. CaWO$_{4}$ is uniaxially anisotropic with the components of the dielectric tensor taking the following form,
\begin{equation}
	\vec{\varepsilon} = \begin{pmatrix}
		\varepsilon_{\perp} & 0 & 0\\
		0 & \varepsilon_{\perp} & 0\\
		0 & 0 & \varepsilon_{||}
	\end{pmatrix}.	
	\label{eq:perm}
\end{equation}
Electromagnetic modes couple to the permittivity tensor with proportion to the electric filling factor (\textit{p}$_{i}$) in the i = x, y, z directions,
\begin{equation}
	p_{i} =	\frac{\iiint_V \varepsilon_{i} E_{i}\cdot E_{i}^{*}dv}{\iiint_V \varepsilon(v) \vec{E} \cdot \vec{E}^{*} dv},
	\label{eq:EFF}
\end{equation}
where $E_i$ is the electric field component inside the sample, and $\vec{E}$ is the total electric field. The integration occurs over the volume of the resonator. For a WGM, the components of $\vec{p}$ sum to near unity, reflecting that the majority of the electromagnetic field is confined within the crystal. WGH modes predominantly couple to the parallel component of the permittivity, $\varepsilon_{||}$, while WGE modes are primarily sensitive to the perpendicular component, $\varepsilon_{\perp}$, though a degree of hybridisation between components is generally present. 
\section{Room-temperature Characterisation}
\begin{table*}[!ht]
    \centering
    \begin{tabular}{c|ccc|ccc}
    \multicolumn{1}{c|}{}&
    \multicolumn{3}{c|}{WGH$_{m,1,1}$}&
    \multicolumn{3}{c}{ WGE$_{m,1,1}$}\\
    \hline
    \hline
    m & Simulation (GHz) & Experiment (GHz) & \% Difference  & Simulation (GHz)& Experiment (GHz) & \% Difference \\
    \hline
$2$&	$3.24221$&	$3.24532$&	$-0.0959$&	$3.35398$&	$3.37555$& $-0.641$\\
$3$&	$4.17384$&	$4.17516$&	$-0.0316$&	$4.39215$&	$4.40504$& $-0.293$\\
$4$&	$5.10694$&	$5.10627$&	$0.0131$&	$5.32309$&	$5.33241$&	$-0.175$\\
$5$&	$6.03444$&	$6.03458$&	$-0.0023$&  $6.21986$&	$6.22772$& $-0.126$\\
$6$&	$6.95423$&	$6.95470$&	$-0.0068$&	$7.09944$&	$7.10453$&	$-0.072$\\
$7$&	$7.86535$&	$7.86657$&	$-0.0155$&	$7.96274$&	$7.96562$&	$-0.036$\\
$8$&	$8.76699$&	$8.767352$&	$-0.0041$&	$8.80933$&	$8.81032$&	$-0.011$\\
$9$&	$9.65747$&	$9.658112$&	$-0.0066$&	$9.64158$&	$9.64073$&	$0.0088$%\\
% $10$&	$10.5320$&	$10.53422$&	$-0.021$&	$10.46312$&	$10.46092$&	$0.021$
    \end{tabular}
    \caption{Room-temperature comparison of measured and simulated WGM frequencies, using $\varepsilon_{||} = 9.029$ and $\varepsilon_{\perp} = 10.761$. For $m=6$ to $m=9$ modes, the mismatch is between $0.004\%$ and $0.072\%$, indicating strong linearity over this domain of mode numbers. Considering the dimensional uncertainty from the crystal radius of $\pm 0.025\%$; conservatively, we estimate the total uncertainty in the extracted permittivity values to be $\pm 0.1\%$.
    Above $m=9$, the mode density increases, such that interactions with cavity modes reduce the precision.}
    %, however the calculation is dominated by dimension uncertainties, conservatively estimated to be 0.1\%, so the simulation precision does not significantly contribute.}
    \label{tab:RT_ID}
\end{table*}
To understand the mode structures within the crystal, we first measured mode frequencies between $2$ and $13$ GHz at room-temperature, with the setup as shown in Fig. \ref{fig:nc_setup}. Several resonances were measured from the evanescent fields that radiate from the crystal surface in the open configuration. The orientation of the probes, with the loops facing vertically or horizontally, enabled preliminary mode classification as quasi-transverse-magnetic (TM) or quasi-transverse-electric (TE), respectively. One probe mount remained fixed at mid-height relative to the crystal, while the other was adjusted incrementally in either the azimuthal ($\phi$) or axial (\textit{z}) direction. For azimuthal scans, the adjustable probe was translated using the lever in steps of $\Delta \phi = 2.5$\textdegree. For axial scans, the probe was initially set just below the base of the crystal and then raised in 5 mm increments. Probe 1 served as a coherent drive, injecting microwave modes synthesised by the Vector Network Analyser (VNA), stabilised by a H-maser reference. Probe 2 passively monitored the evolution of the field, characterising the coupling efficiency. At each probe configuration, the transmission data (scattering parameter, $S_{21}$) for a frequency sweep across the resonance were measured. Counting the number of $S_{21}$ field variations along those axes allowed for clear confirmation of the mode identity, as seen in Fig. \ref{fig:hz_field}, where high transmission corresponded to the antinodes of a WGM; the result of coupling strength being maximised when the adjustable probe is located at an antinode. Following the experimental identification of resonances, the sample was relocated into a copper cavity to mitigate external electromagnetic interference and radiative loss, and establish defined boundary conditions.   At room-temperature, the dielectric sample's dimensions were measured to be $r_{d} =  20.00 \pm 0.005$ mm and $h_{d} = 39.00 \pm 0.01$ mm for radius and height, respectively. The crystal was mounted on a sapphire post ($r_{p}=2.50\,$mm, $h_{p}=10.0\,$mm from copper base to crystal) through a hole drilled in the bottom face of the crystal, and clamped inside the oxygen\hyp free copper cavity ($r_{c}=30.0\,$mm, $h_{c}=60.0\,$mm). For WGMs, introducing the cavity-wall boundary condition primarily affects the evanescent field outside the crystal and therefore only weakly perturbs the resonance frequency, while suppressing radiation-loss channels and improving the $Q$ factor. As a result, WGMs remained clearly identifiable in the measured spectrum, even in the presence of a denser set of cavity modes. This distinction was possible because the WGMs exhibited significantly higher $Q$ factors than the coexisting cavity resonances. \\\\
Using finite element modelling software, COMSOL, the simulated eigenfrequency mode solutions were matched to the measured resonant frequencies by parametrically sweeping across input permittivity values. Fig. \ref{fig:hz_field} shows an example of the strong concordance between the numerical simulation results and the experimentally measured electromagnetic field distributions, thereby validating the mode assignment and enabling the accurate determination of the associated electric filling factors. The results for the fundamental WGM families are summarised in Fig. \ref{fig:m_vs}. Fig.~\ref{fig:m_vs}A presents the simulated frequency solutions for the fundamental WGM families, with a comparison to measured values summarised in Table~\ref{tab:RT_ID}. Optimal agreement between measured WGM frequencies and simulation was attained for $\varepsilon^{295\, K}_{||} = 9.029 \pm 0.009$ and $\varepsilon^{295\, K}_{\perp} = 10.761 \pm 0.011$. The component of the dielectric tensor aligned with the crystal $c$-axis, $\varepsilon_{||}$, agrees well with values previously reported at MHz frequencies, remaining within the associated uncertainties. However, the discrepancy in measured values of this work and all other literature is notably different ranging from $4.76\%$ to $8.03\%$, with the greatest variability between the reported values for $\varepsilon_{\perp}$. While frequency dispersion in the permittivity may contribute to these differences, it is unlikely to be the dominant cause at microwave frequencies, since the first optical phonon modes that significantly affect permittivity lie in the THz range~\cite{HAYASAKA2007386}.\\\\
Fig. \ref{fig:m_vs}B shows the simulated electric filling factors of the excited microwave modes calculated from Eq. \ref{eq:EFF}. These two graphs confirm WGM-like behaviour at m$=6$ and above. Below this point, Fig. \ref{fig:m_vs}B shows the resonances are more dispersive, and therefore less reliable for determining dielectric tensor values. This is also reflected in the frequencies in Fig. \ref{fig:m_vs}A, where the FSR diverges as m $\xrightarrow{} 0$. This cutoff point dividing the high-frequency WGM region of interest and low-frequency dispersive region is marked by a grey line. In the high-frequency regime, the mode field confinement is sufficient to establish the dielectric permittivity tensor values and the associated dielectric losses, which will be discussed in more detail later. The FSR of the WGMs was observed to be nearly constant for m$\ge6$ in Fig. \ref{fig:m_vs}A, and a linear fit yields FSR values of 0.830 GHz for WGE modes and 0.897 GHz for WGH modes. For azimuthal mode numbers of $m \ge 10$, hybridisation between the fundamental WGE and WGH mode families becomes significant. This hybridisation introduces complexity into mode classification, necessitating careful consideration when identifying and assigning WGM types in this regime.
\section{Low-Temperature Characterisation}
\begin{figure}[t!]
		\includegraphics[width=0.8\columnwidth]{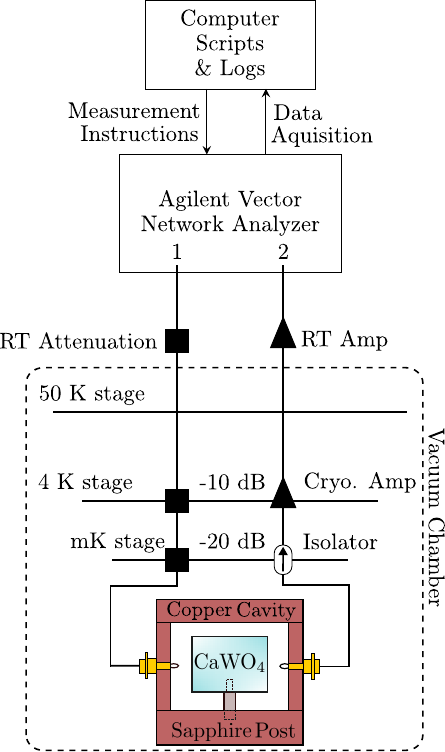}
			\caption{Schematic of resonator setup in dilution refrigerator.}
			\label{fig:fridge_setup}
\end{figure}

\begin{table}[!ht]
	\centering
 \begin{tabular}{ccccc}
		\hline
		$\varepsilon_{||}$ &$\varepsilon_{\perp}$&Freq.&Temp. (K)&Ref.\\
		\hline
		
		 $8.7937$&$10.4400 $& GHz & $4$& *\\
        $8.7967$&$10.4417$& GHz &$25$& * \\
        $8.8001$&$10.4474$& GHz &$50$& * \\
        $8.8164$&$10.4703$& GHz & $100$& *\\
        $8.8472$&$10.5166$& GHz & $150$& *\\
        $8.8923$&$10.5806$& GHz & $200$& *\\
        $8.9583$&$10.6713$& GHz & $250$& *\\
        $9.0290$&$10.7611$& GHz &  $295$& *\\
	$9.1 \pm 0.4$& $11.3 \pm 0.4$ &$1-40$ MHz&$293.15$&\cite{Thorp1975}\\
        $9.5 \pm 0.2$& $11.7 \pm 0.1$ &$1.59$ kHz&$297.65$&\cite{BowerandFang,DCofScheelite}
	\end{tabular}
	\caption{Components of the dielectric tensor for undoped CaWO$_{4}$. Values derived in this work$^*$  carry an error of $\pm 0.1\%$.}
	\label{tbl:permreport}
\end{table}
\begin{figure}[t!]
			\includegraphics[width=1.0\columnwidth]{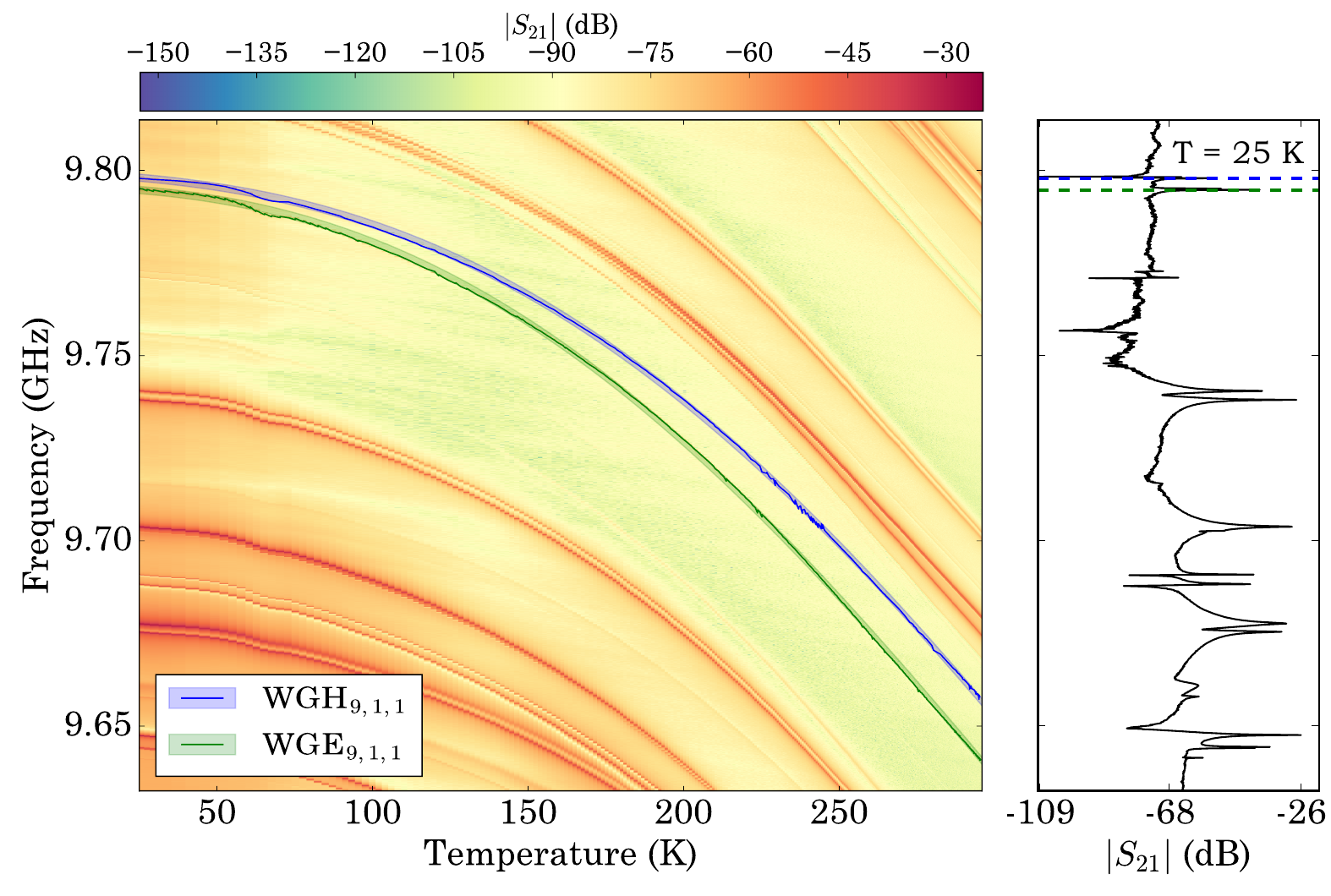}
			\caption{The WGH$_{9,1,1}$ (blue) and WGE$_{9,1,1}$ (green) were used to match eigenfrequency solutions from simulations to the experimental values. The solid line follows the measured resonance, and the lighter bands represent theoretical predictions and $0.02\%$ error encapsulating residuals. Error here is associated with the combination of simulation error and random fluctuations in measured frequencies. The spectrum on the right illustrates a temperature `slice' of the colour density plot at $25$ K with the WGH$_{9,1,1}$ (blue) and WGE$_{9,1,1}$ (green) resonances marked with dashed lines. }
			\label{fig:t_vs_f}
\end{figure}
\begin{figure}[h!]
    \centerline{\begin{minipage}{\columnwidth}
			\centering
			\includegraphics[width=0.9\columnwidth]{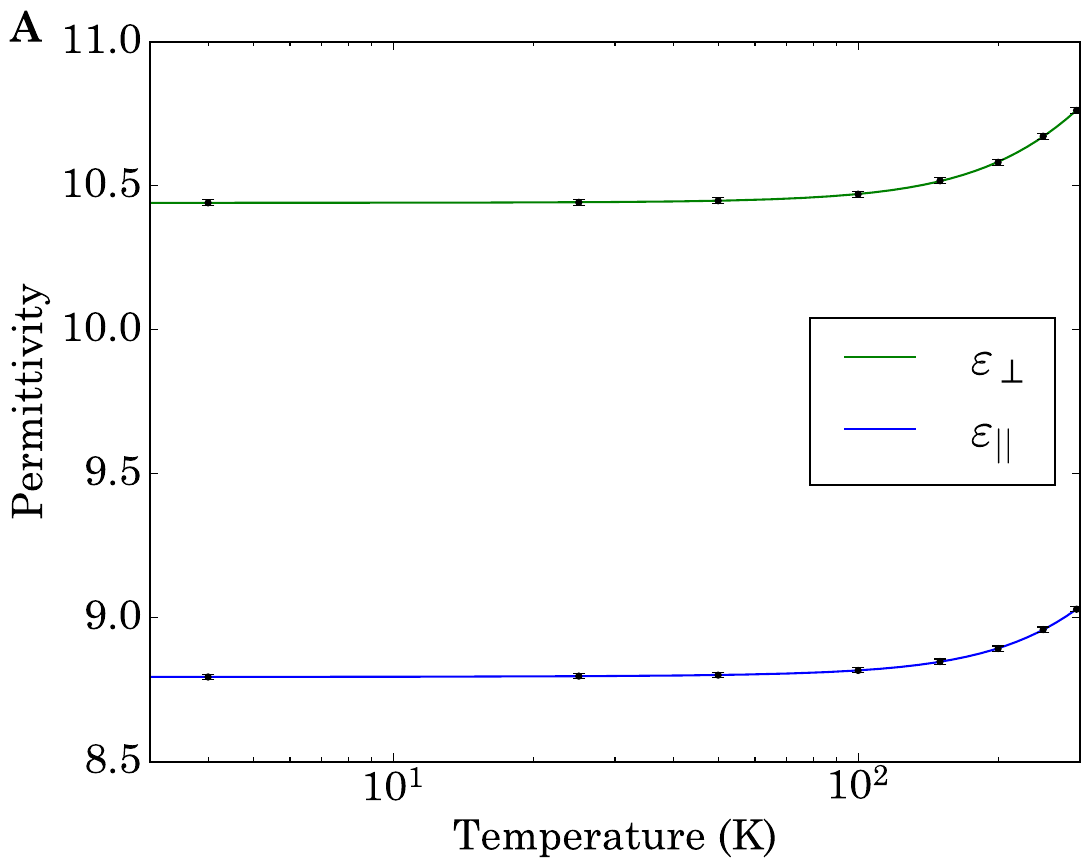}
			\includegraphics[width=0.9\columnwidth]{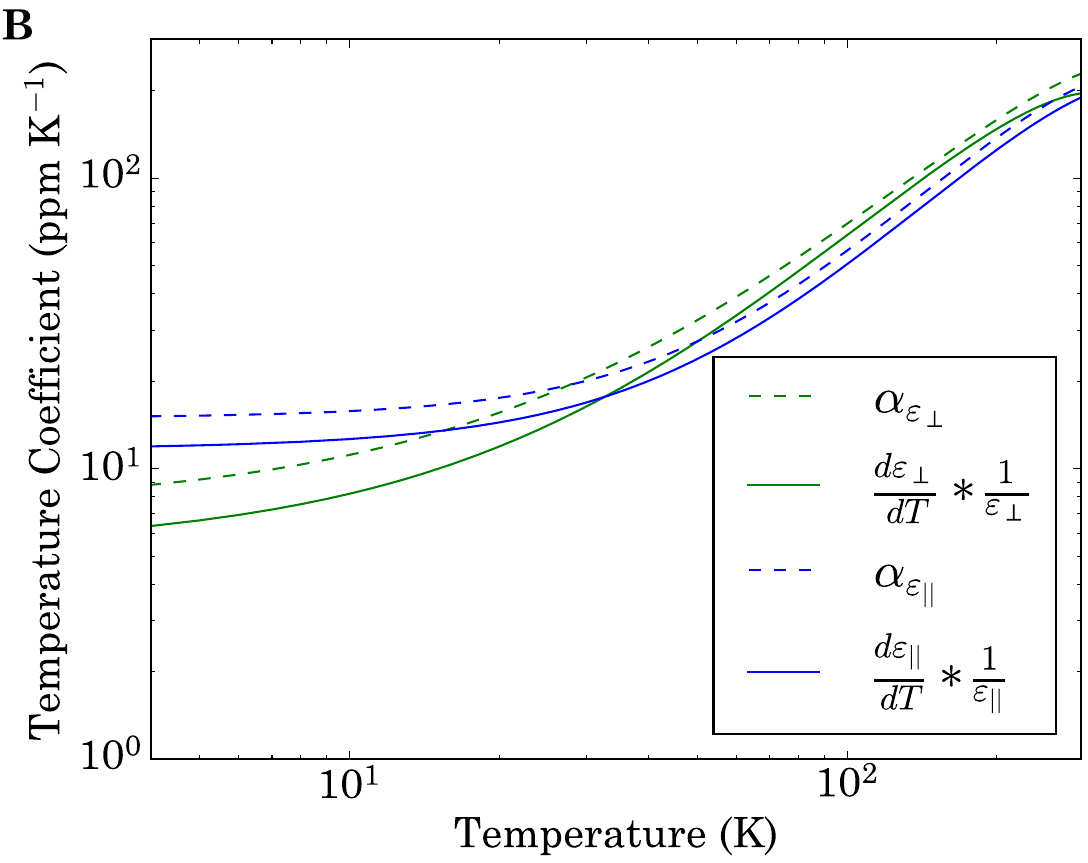}
			\caption{\textbf{A} The parallel (blue) and perpendicular (green) components of the dielectric tensor as a function of temperature. \textbf{B} The temperature coefficients of permittivity ($\alpha_{\epsilon_i}$) for CaWO$_{4}$ as a function of temperature, calculated by the derivative of the permittivity (solid lines) and Eq. \ref{eq:alpha} (dashed lines).}
			\label{fig:t_vs_perm}
	\end{minipage}}
\end{figure}
The temperature dependence of a WGM resonance frequency, $f(T)$, can be decomposed into two contributions: (i) the change in dielectric permittivity with temperature and (ii) the change in the crystal dimensions due to thermal contraction. In this work, both effects are included. The permittivity variation was obtained using two complementary approaches: numerical simulation and an analytical calculation. In this section, we analyse the measured frequency shifts, correcting for geometric contraction, to extract the temperature dependence of the permittivity.\\\\
The relationship between linear coefficients of temperature ($\alpha_{\chi}$) and the associated parameter ($\chi$) can be written generically as;
\begin{equation}
    \chi(T) =\chi(T_{ref}) + \chi(T_{ref})\int \limits_{T_{ref}}^{T}\alpha_{\chi}(\tau)\, d\tau \,, 
    \label{eq:alpha_general}
\end{equation}
where $\chi$ is a placeholder for either dimensional parameters, length ($L = h_{d}$) and diameter ($D = 2r_{d}$), or $\vec{\varepsilon}$. $\tau$ is a temperature dummy variable of integration, and $T_{ref}$ is the reference temperature at which the dimensions and permittivity are measured (in this case $T_{ref}=$ 295 K). Here, $\alpha_{D}$ and $\alpha_{L}$ are the temperature coefficients of linear expansion for the crystal diameter and length, respectively, and $\alpha_{\varepsilon_{\perp}}$ and $\alpha_{\varepsilon_{\parallel}}$ are the temperature coefficients of the permittivity perpendicular and parallel to the crystal c-axis, respectively (see Eq.~\ref{eq:alpha}). We used the thermal expansion parameters measured by Senyshyn \textit{et al.} \cite{PhysRevB.70.214306}. Eq. ~\ref{eq:alpha_general} was then implemented to update the crystal dimensions as a function of temperature as it was cooled from $T_{\mathrm{ref}}$ = 295 K to cryogenic temperatures.\\\\
The frequency shifts of the WGMs were measured as the sample was cooled from room-temperature down to 4 K and subsequently $100$ mK using a dilution refrigerator, with the setup shown in Fig. \ref{fig:fridge_setup}. The CaWO$_4$ dielectric crystal was calculated to contract to $r_{d} = 19.97 \pm 0.005$\,mm, $h_{d} = 38.88 \pm 0.01$\,mm at low temperatures. Input RF signals were attenuated to avoid unwanted heating and non-linear effects, and output signals were amplified using cryogenic low-noise amplifiers (LNA) to enhance SNR. The loop probes were oriented at $45^{\circ}$ to excite WGMs of both mode families. With particular focus on two WGMs of different mode families close together in frequency space (WGH$_{9,1,1}$ and WGE$_{9,1,1}$), we determined the frequency shift with temperature ($\frac{\partial f}{\partial T}$) as shown in Fig. \ref{fig:t_vs_f}. The advantage of these two select modes is that they both exhibit high filling factors in the parallel and perpendicular directions, respectively, as shown in Fig. \ref{fig:m_vs}B. This allows for a clear separation of effects on frequency due to changes in the corresponding components of $\vec{\varepsilon}$. These modes shall henceforth be referred to as calibration modes. Fig. \ref{fig:t_vs_perm}A illustrates the temperature dependence of the $\vec{\varepsilon}$ of $\text{CaWO}_4$, as derived by iteratively fitting the resonant frequencies obtained from numerical simulations to the experimental measurements. The results for several temperature points are tabulated in Table \ref{tbl:permreport}.\\\\
In order to conduct the analytical evaluation of the dielectric permittivity change with temperature, we parameterised the contributions of the dielectric permittivity change and dimensional change to the frequency shift with the following relation:
\begin{equation}
	\frac{1}{f} \frac{\partial f}{\partial T} = -\frac{1}{2} (p_{\varepsilon_{\perp}} \alpha_{\varepsilon_{\perp}} +p_{\varepsilon_{||}} \alpha_{\varepsilon_{||}})-p_{D} \alpha_{D}-p_{L}\alpha_{L}.
    \label{eq:alpha}
\end{equation}
Here, the filling factors ($p_{\chi}$) weight the contributions of these two effects, and are defined via the following incremental frequency rules \cite{WGMDCDETERMINATION},
\begin{align}
	\begin{split}
		p_{\varepsilon_{\perp}} &= 2 \Big| \frac{\partial f}{\partial \varepsilon_{\perp}} \Big| \frac{\varepsilon_{\perp}}{f},\\
		p_{\varepsilon_{||}} &= 2 \Big| \frac{\partial f}{\partial \varepsilon_{||}} \Big| \frac{\varepsilon_{||}}{f},\\
		p_{D} &= \Big| \frac{\partial f}{\partial D} \Big| \frac{D}{f},\\
		p_{L} &= \Big| \frac{\partial f}{\partial L} \Big| \frac{L}{f}.
	\end{split}
\end{align}
By applying Eq. \ref{eq:alpha} to the calibration modes and solving the simultaneous equations, we can recover the temperature coefficients of the dielectric permittivity ($\alpha_{\varepsilon_{\perp}}$, $\alpha_{\varepsilon_{||}}$). Fig. \ref{fig:t_vs_perm}B shows the temperature coefficients as computed from the normalised derivative of $\vec{\varepsilon}$ (solid lines) and is compared to the empirical result calculated from Eq. \ref{eq:alpha} (dashed lines). This comparison is valuable for checking the self-consistency of simulation data and quantifying the various sources of error. Error in the permittivity stems from a combination of mesh error in simulation, measurement error of the frequencies using the VNA, calibration of the thermistors in the cryostat, and uncertainty in the crystal dimensions. Conservatively, these errors total to approximately $0.1\%$, with uncertainty in $r_{d}$ constituting the dominant contribution to overall experimental error. This corresponds to frequency shifts in simulation of $\mathcal{O}{(10^6)}$ Hz, which is orders of magnitude greater than the linewidth of most WGM, especially at low temperatures. Accuracy may be preserved by including a third mode far from the calibration modes already chosen to introduce a third degree of freedom that would constrain the value of $r_{d}$. Thus, the precision of the permittivity can be meaningfully increased in future studies using WGM analysis. 
\section{Loss Mechanisms}
\begin{figure}[t!]
	\centerline{\begin{minipage}{\linewidth}
			\centering
			\includegraphics[width=\textwidth]{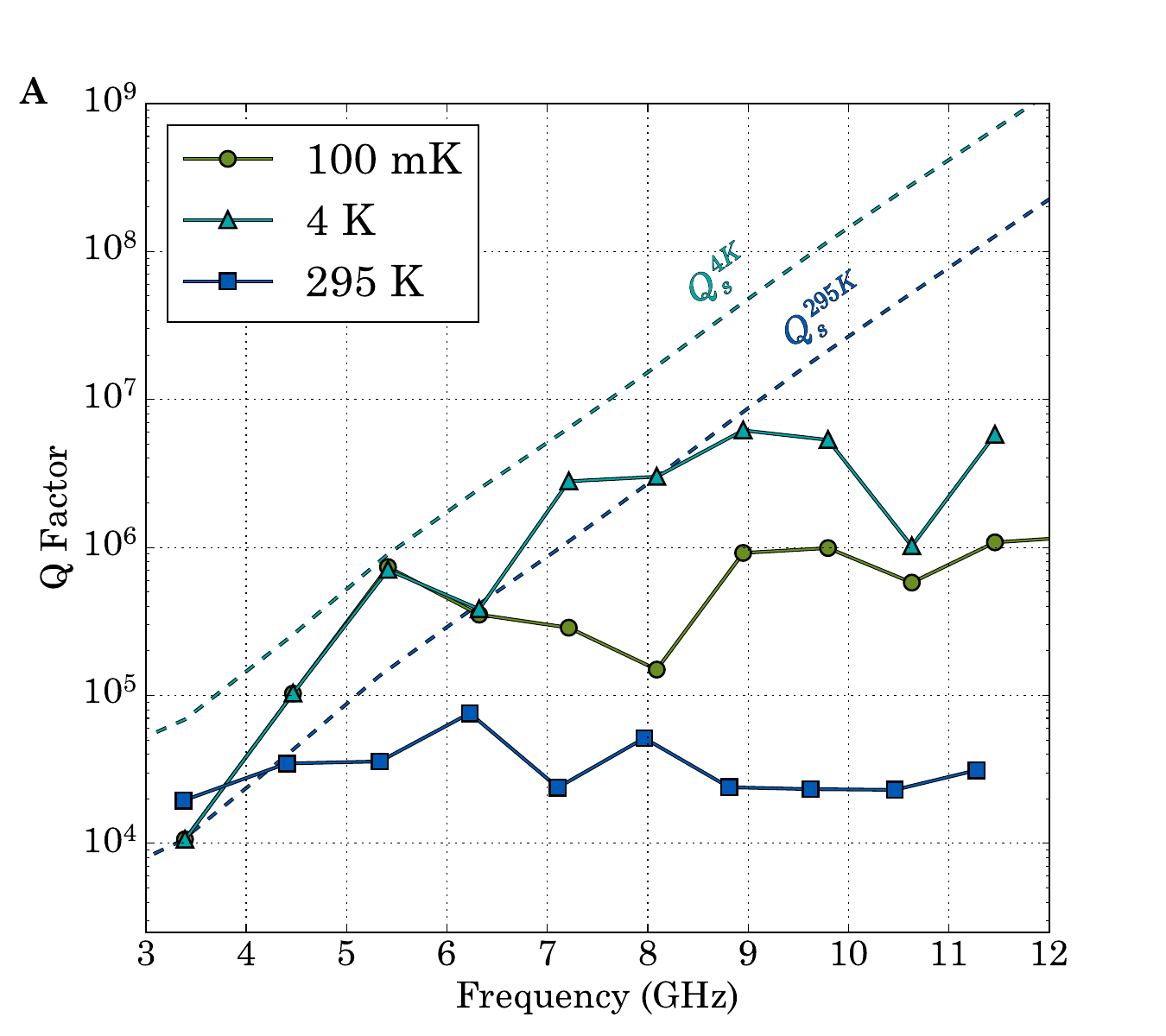}
			\includegraphics[width=\textwidth]{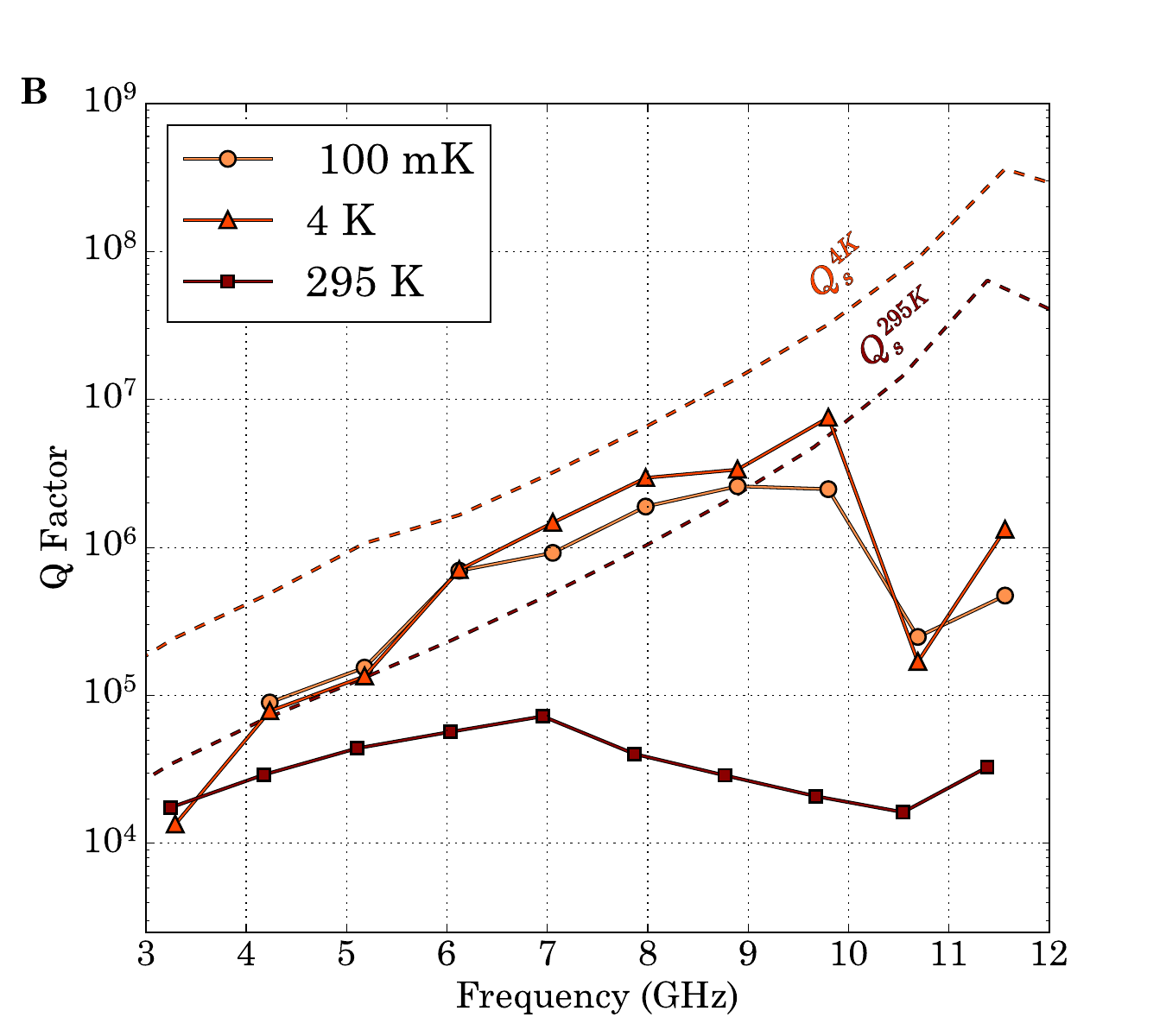}
			\caption{The unloaded $Q$-factors of the fundamental whispering gallery mode families (\textbf{A}: WGE$_{m,1,1}$, \textbf{B}: WGH$_{m,1,1}$). The dashed lines show the upper bounds of the unloaded $Q$ due to the surface resistance of the copper shield with $R_{s}^{295 K}  = 0.13 \pm 0.03 \, \Omega$ and $R_{s}^{4 K}  = 0.018 \pm 0.004 \, \Omega$.  }
			\label{fig:qfactors}
	\end{minipage}}
\end{figure}

Loss mechanisms were characterised for the shielded sample, at room-temperature (295 K), liquid-helium temperature (4 K), and superfluid helium temperature (100 mK).  Fig. \ref{fig:qfactors} displays unloaded $Q$ factors for the two mode families, determined from fitting to the undercoupled \cite{coupling} Fano resonances of the $S_{21}$ transfer functions. We can infer the effect of loss mechanisms in different frequency regions. At cryogenic temperatures, we see $Q$ factors increase with frequency until they plateau, suggesting that $Q$ factors are limited proportionally with cavity wall losses until the WGMs are confined enough to be dominated instead by loss mechanisms within the crystal. The geometry factor quantifies the loss contribution due to the copper cavity walls as
\begin{equation}
G =	\omega \frac{\iiint_V \mu_{0} \vec{H}\cdot \vec{H}^{*} dv}{\iint_S \vec{H}_{t} \cdot \vec{H}^{*}_{t} ds},
	\label{eq:GF}	
\end{equation}
where the total magnetic field is given in the numerator by integrating $\vec{H}$ over the volume of the resonator, and $\vec{H}_{t}$ is the tangential magnetic field integrated over the inner surface of the cavity. The geometry factors in units of Ohms ($\Omega$) are tabulated in Table \ref{tbl:Greport}.
For any select mode, the effect of the cavity wall surface resistance ($R_s)$ can be calculated with
\begin{equation}
	 Q_{s} = G/R_s .
	\label{eq:Rs}	
\end{equation}
Careful characterisation of these parameters allows us to determine the dielectric loss tangent,
\begin{equation}
	Q^{-1} = p_{\perp} tan\delta_{\perp} +p_{||} tan\delta_{||} + R_{s}/G + Q_{r}^{-1},
	\label{eq:losstangent}	
\end{equation}
where $Q$ is the unloaded \textit{Q} factor and $Q_{r}^{-1}$ accounts for the radiative losses, which are negligible in the case of an undercoupled shielded dielectric resonator. Coupling parameters were calculated from the probe $1$ and $2$ reflection data ($S_{11}$ and $S_{22}$) and were set at room temperature to be $\mathcal{O}{(10^{-4}-10^{-3})}$ to ensure they remained under-coupled for the full temperature range. 
$R_{s}$ was determined by measuring $Q$ factors of the resonant modes in the empty oxygen-free cavity at room-temperature. The $R_{s}$ at cryogenic temperature was derived from residual resistivity ratio (RRR) values \cite{Calatroni2020MaterialsPropertiesTC,1062352}. At low frequency, the $Q$ factors are dominated by the loss contribution of the $G/R_{s}$ term, as indicated by the little to no change in $Q$ between $4$ K and $100$ mK where $R_s$ bottoms out for the case of non-ideal conductors.
\begin{table}[!t]
	\centering
	\begin{tabular}{ccccc}
		\hline
       &\multicolumn{2}{c}{WGH} & \multicolumn{2}{c}{WGE} \\
       \hline
		$m$ &$4$ K&$295$ K&$4$ K&$295$ K\\
		\hline
		$1$&$1844$&1827&$515$&$520$\\
        $2$&$4402$&$4288$&$1241$&$1305$\\
        $3$&$8808$&$8640$&$4560$&$4886$\\
        $4$&$19257$&$15785$&$16005$&$16922$\\
        $5$&$29845$&$29393$&$44477$&$46632$\\
        $6$&$57898$&$57305$&$114875$&$121463$\\
        $7$&$117811$&$117285$&$303013$&$325099$\\
        $8$&$251927$&$253850$&$813758$&$882175$\\
        $9$&$580987$&$600478$&$2104466$&$2277235$\\
        $10$&$1614229$&$1793383$&$5116668$&$5521704$\\
        $11$&$6517959$&$8009537$&$12129278$&$13136379$\\
        $12$&$4017406$&$3877068$&$28690955$&$31091121$
    \end{tabular}
	\caption{Geometry factors ($\Omega$) from numerical simulation at 4 K and 295 K.}
	\label{tbl:Greport}
\end{table}
The change in trend for frequencies above $6$ GHz, corresponding to $m \geq 6 $, indicates dielectric loss prevails in room-temperature and liquid helium temperature conditions. However, at superfluid helium temperature, $Q$ factors degrade from their $4$ K value, a phenomenon that may be attributed to spins from impurity ions settling into their ground state and becoming the dominant zero-field loss mechanism for resonant WGM frequencies lying in the vicinity of the spin transition energy \cite{Creedon11,Goryachev14}. Exploration of spin impurities will be conducted in future studies. Therefore, the loss tangent is compared only between $295$ K and $4$ K in the frequency range where dielectric loss dominates. 
\begin{figure}[t!]
	\centerline{\begin{minipage}{\linewidth}
			\centering
			\includegraphics[width=\textwidth]{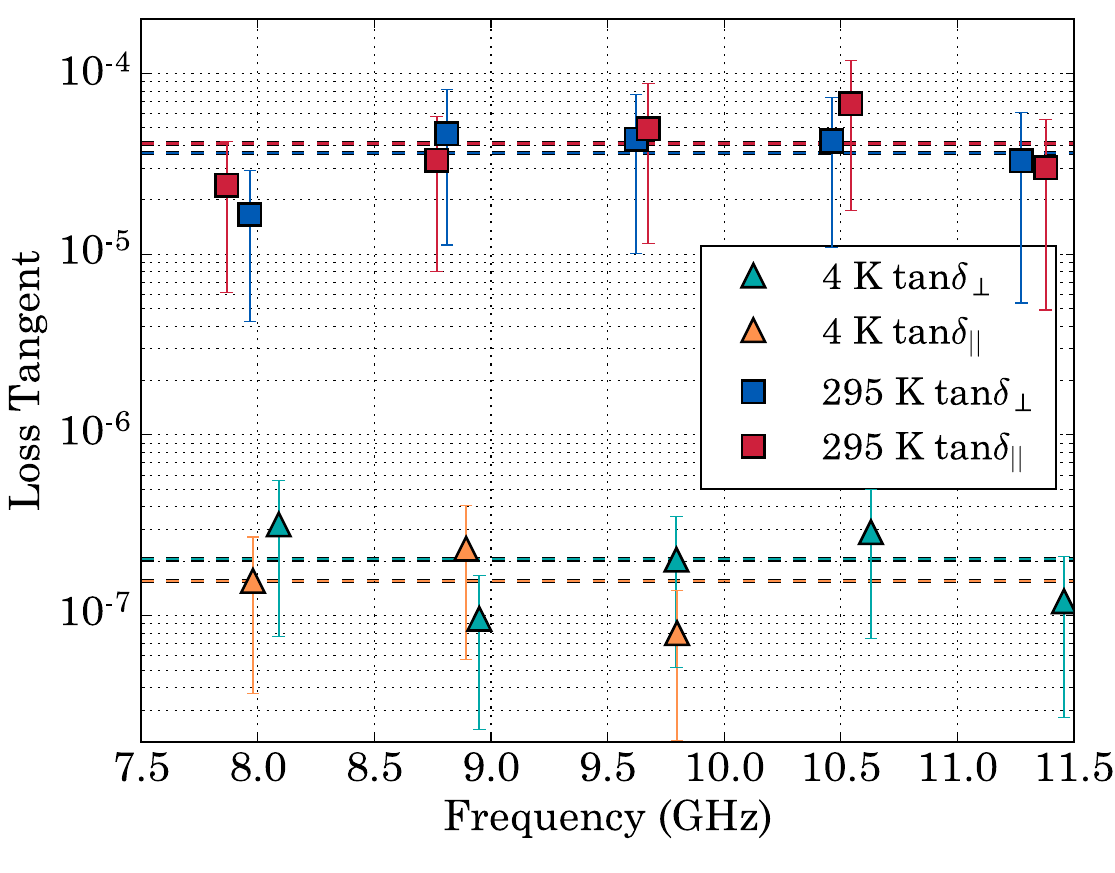}
			\caption{Components of the loss tangent of CaWO$_{4}$ at room-temperature (squares), and liquid helium temperature (triangles). The mean values are indicated by dashed lines:  tan$\delta_{||}^{295 K} = 4.1\pm 1.4 \times 10^{-5}$ (red), tan$\delta_{\perp}^{295 K} = 3.64\pm 0.92 \times 10^{-5}$ (blue), tan$\delta_{||}^{4 K} = 1.56\pm 0.52 \times 10^{-7}$ (orange), tan$\delta_{\perp}^{4 K}= 2.05\pm 0.79 \times 10^{-7}$ (green).}
			\label{fig:losstan}
	\end{minipage}}
\end{figure}
To calculate the components of the loss tangent for this uniaxial sample, a mode from each fundamental WGM family was selected, and the pair of simultaneous equations was solved, see Eq. \ref{eq:losstangent}. The components of the loss tangent of CaWO$_{4}$ at room-temperature and liquid helium temperature are plotted in Fig. \ref{fig:losstan}. The room-temperature value  was found to be tan$\delta^{295\,K}_{||} = 4.1\pm 1.4 \times 10^{-5}$, tan$\delta^{295\,K}_{\perp} = 3.64\pm 0.92 \times 10^{-5}$. At liquid-helium temperature the loss tangent was tan$\delta^{4\,K}_{||} = 2.05\pm 0.79 \times 10^{-7}$, tan$\delta^{4\,K}_{\perp} = 1.56\pm 0.52 \times 10^{-7}$. Previous investigations on a different CaWO$_{4}$ sample indicate that these values are an order of magnitude worse than expected \cite{Hartman24}. Results plotted in Fig. \ref{fig:qfactors} suggest a loss peak around 10.5 GHz at 4 K, most likely due to an unknown impurity. The effect is broad enough to degrade the $Q$ factor of both WGE and WGH modes between $9$ to $12$ GHz. This effect highlights the importance of attaining high-purity materials for precision and quantum measurements. Future work will utilise WGM spectroscopy at mK temperature \cite{Farr13} to understand the impurities in this crystal. WGM investigations in other crystals such as SrLaAlO$_{4}$ demonstrate that by polarising all spins at a high magnetic field, the magnetic loss can be made negligible \cite{Hosain18}.\\\\
\section{Conclusion}
We presented a study employing WGM analysis to evaluate the dielectric permittivity of single-crystal CaWO$_{4}$ from $295$ K to $4$ K, and determined the relative permittivity tensor at GHz frequencies. These values are important for the future design of microwave circuits that will utilise this crystal. The loss mechanisms associated with the WGM mode analysis and the components of the loss tangent are also reported, and for this sample, were limited by an unknown paramagnetic impurity around 10.5 GHz.\\\\
The loss tangent of CaWO$_4$, while approximately an order of magnitude higher than that of sapphire, is comparable to that of YAG and notably lower than those of SrLaAlO$_4$, rutile, and quartz \cite{JerzyKrupka_1999}. These results highlight CaWO$_4$ as a promising candidate for emerging sensing applications and provide valuable data on the temperature dependence of its dielectric permittivity, contributing to the broader scientific understanding of this material.

\subsection*{Acknowledgement}
This research was supported by the ARC Centre of Excellence for Dark Matter Particle Physics (CE200100008).

\end{document}